\documentclass{article}
\usepackage{arxiv}

\usepackage[utf8]{inputenc} 
\usepackage[T1]{fontenc}    
\usepackage[hidelinks]{hyperref}       
\usepackage{url}            
\usepackage{booktabs}       
\usepackage{amsfonts}       
\usepackage{nicefrac}       
\usepackage{microtype}      
\usepackage{lipsum}		
\usepackage{graphicx}
\usepackage[square,numbers]{natbib}
\usepackage{doi}
\usepackage{caption}
\usepackage{subcaption}
\usepackage{amsmath}

\newcommand{\mean}[1]{\overline{#1}\,}
   \newcommand{\logmean}[1]{\overline{#1}^{\text{log}}}
   \newcommand{\gmean}[1]{\overline{#1}^{G}}
   \newcommand{\hmean}[1]{\overline{#1}^{H}}
   \newcommand{\diff}{\delta}

\usepackage{bm}
\renewcommand{\vec}[1]{\bm{#1}}

\title{Asymptotically entropy-conservative and kinetic-energy preserving numerical fluxes for compressible Euler equations}

\author{ \href{https://orcid.org/0000-0002-6518-3114}{\includegraphics[scale=0.06]{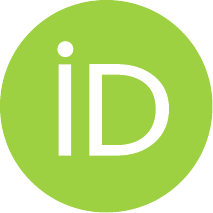}\hspace{1mm} Carlo {De~Michele}}\\
	Dipartimento di Ingegneria Industriale\\
	Universit\`a di Napoli ``Federico II''\\
	Napoli, Italy \\
	\texttt{carlo.demichele2@uninait} \\
	\And
	\href{https://orcid.org/0000-0003-4943-9551}{\includegraphics[scale=0.06]{orcid.pdf}\hspace{1mm}Gennaro Coppola} \\
	Dipartimento di Ingegneria Industriale\\
	Universit\`a di Napoli ``Federico II''\\
	Napoli, Italy \\
	\texttt{gcoppola@uninait} \\
}

\renewcommand{\shorttitle}{AEC and KEP numerical fluxes for compressible Euler equations}

\hypersetup{
pdftitle={Asymptotically entropy-conservative and kinetic-energy preserving numerical fluxes for compressible Euler equations},
pdfsubject={physics.flu-dyn},
pdfauthor={Carlo De~Michele, Gennaro Coppola}
pdfkeywords={Compressible flow, Finite-volume, Entropy conservation, Pressure equilibrium preservation},
}

\begin{document}
\maketitle

\begin{abstract}
This paper proposes a hierarchy of numerical fluxes for the compressible flow equations which are kinetic-energy and pressure equilibrium preserving and asymptotically entropy conservative, i.e., they are able to arbitrarily reduce the numerical error on entropy production due to the spatial discretization. The fluxes are based on the use of the harmonic mean for internal energy and only use algebraic operations,  making them  less computationally expensive than the entropy-conserving fluxes based on the logarithmic mean.
The use of the geometric mean is also explored and identified to be well-suited to reduce errors on entropy evolution. Results of numerical tests confirmed the theoretical predictions and the entropy-conserving capabilities of a selection of schemes have been compared.
\end{abstract}

\keywords{Compressible flow \and Finite-volume \and Entropy conservation \and  Pressure equilibrium preservation}

\section{Introduction}\label{Introduction}
It is well known that, at high Reynolds numbers, even in shock-free conditions, numerical simulations of compressible flows are strongly affected by nonlinear instabilities,
mainly arising from the spatial discretization of the convective terms in the Euler equations.
Over the past years, many strategies have been implemented to alleviate this phenomenon, the most pursued one being the design of numerical discretizations able to guarantee a correct balance of suitably selected induced secondary quantities~\citep{Coppola2019b}.
Kinetic Energy Preserving (KEP) methods, which are able to discretely reproduce the correct kinetic-energy balance due to
convective terms, are probably the most popular among them and have received much attention in recent years. They have shown increased robustness in under-resolved turbulent simulations and are routinely employed nowadays.
A quite general class of KEP schemes for Finite-Difference (FD) discretizations has been recently proposed~\citep{Coppola2019} and the corresponding fluxes have been characterized~\citep{Veldman2019,Coppola2023}.

In addition to the case of kinetic energy, the discrete preservation of the entropy balance has been an important topic of research as well~\cite{Tadmor1987,Ismail2009,Ranocha2018}.
However, in contrast to KEP schemes, which can be formulated as classical FD discretizations of the divergence and advective forms of the convective terms (and are associated with algebraic fluxes), Entropy Conservative (EC) schemes are
almost exclusively introduced in the context of Finite Volume (FV) methods.
As such, they are formulated by directly specifying the numerical fluxes, which typically are nonlinear and
require the evaluation of costly transcendental functions, with a non-negligible increase in computational cost when compared to classical FD  discretizations. 
Moreover, the most popular EC fluxes are based on the logarithmic mean~\cite{Ismail2009}, which needs a local treatment to avoid division by zero, leading to a less straightforward implementation.

In this note, we develop a new class of Asymptotically Entropy Conservative (AEC) schemes suitable for shock-free regions of compressible flows. They are based on 
economic algebraic fluxes and, while retaining the classical KEP property, provide a hierarchy of approximations with increasingly accurate entropy-conservative properties. Moreover, in contrast to existing asymptotic expansions approximating exact EC fluxes~\cite{Tamaki2022}, the proposed approach is able to retain the Pressure Equilibrium Preserving (PEP) property~\cite{Shima2021} at each order of approximation. 
The method is based on a direct specification of the fluxes for mass, momentum, and internal energy, although they can be equivalently reformulated by specifying the total energy flux.
One of its main features is that it is based on the harmonic mean for the internal energy convective flux, which is shown to provide a better approximation of the entropy-conservative flux, as compared to the classical schemes based on the arithmetic mean. 

\section{Existing KEP fluxes with entropy conservation properties}
Our investigation starts with the analysis of some existing schemes for the compressible Euler equations. 
The main properties we seek are: (i) KEP, i.e. the discretized convective terms in mass and momentum equations induce a conservative structure of the convective term in the kinetic-energy balance~\cite{Coppola2019,DeMichele2023}; (ii) EC, i.e. the discretization of mass and energy equations induce a conservative structure of the convective term in the entropy equation~\cite{Tadmor1987,Ranocha2018}; (iii) PEP, i.e. the numerical method is able to preserve the property that an initial condition with constant distribution of pressure $p$ and velocity $u$ induces time derivatives $\partial_tp$ and $\partial_t u$ everywhere zero: the solution evolves as a density wave~\cite{Shima2021,Ranocha2021}.
In the context of FV formulations, the theory of entropy variables gives sufficient conditions for the set of numerical fluxes to reproduce the EC property~\cite{Tadmor1987}. Similar necessary and sufficient conditions have been derived also for the KEP property, at least for two-point fluxes~\cite{Jameson2008b,Veldman2019}, and for the PEP property~\cite{Ranocha2021}.

We will work in a semidiscretized framework, in which the time derivatives of the conserved variables are driven by a difference of convective and pressure numerical fluxes at adjacent faces. We will illustrate the method with reference to
second-order (two-point) fluxes; the corresponding high-order extension can be constructed by adopting the approach described in \ref{sec:high_order}.

Among the various sets of fluxes proposed in past years, we firstly consider here the flux of Ranocha~\cite{Ranocha2021}, which satisfies KEP, EC and PEP properties and can be expressed as fluxes for mass, momentum and total energy as:
\begin{equation}\label{eq:Ranocha_Flux_Etot}
\mathcal{F}_{\rho}= \overline{\rho}^{\text{log}}\,\overline{u},\qquad\qquad
\mathcal{F}_{\rho u} =\mathcal{F}_{\rho}\,\overline{u}+\overline{p},\qquad\qquad
\mathcal{F}_{\rho E} = \dfrac{1}{2}\mathcal{F}_{\rho}u_iu_{i+1}+\mathcal{F}_{\rho}\,\left[\overline{\left(1/e\right)}^{\text{log}}\right]^{-1} + \overline{\overline{\left(p,u\right)}},
\end{equation}
where $\overline{\phi}=\left(\phi_i+\phi_{i+1}\right)/2$ is the arithmetic mean, $\overline{\phi}^{\text{log}}=\left(\phi_{i+1}-\phi_i\right)/\left(\log(\phi_{i+1})-\log(\phi_i)\right)$ is the logarithmic mean and $\overline{\overline{\left(\phi,\psi\right)}}=\left(\phi_i\psi_{i+1}+\phi_{i+1}\psi_i\right)/2$ is the product mean.
In Eq.~\eqref{eq:Ranocha_Flux_Etot} the flux is expressed in a one-dimensional setting, the three-dimensional extension being easily obtained by adding the componentwise contributions. The usual meaning of the symbols is adopted: $\rho,u$ and $p$ are the density, velocity and pressure, respectively, whereas $e$ and $E$ are the internal and total energy per unit mass, linked by $E = e+u^2/2$. Perfect gas model will be assumed, for which $p=\left(\gamma -1\right)\rho e$, where $\gamma = 1.4$ is the ratio of specific heats. The physical entropy is given by $s = \log(p/\rho^\gamma)$.

The KEP property of the flux in Eq.~\eqref{eq:Ranocha_Flux_Etot} is evident by inspecting the convective term in the momentum flux, which is given by the product between the mass flux $\mathcal{F}_{\rho}$ and the arithmetic average of velocity $\overline{u}$.
This is indeed the necessary and sufficient condition for (second-order, two-point) fluxes to be KEP, even on non-uniform or non-Cartesian meshes~\cite{Veldman2019}. 
One of the key results of recent analyses of the KEP schemes for compressible flows is that, when a KEP scheme is adopted, kinetic energy is conserved globally and locally, with a numerical flux given by $\mathcal{F}_{\rho}u_iu_{i+1}/2$ in the second-order case~\cite{Coppola2023}.
In the expression for $\mathcal{F}_{\rho E}$ in Eq.~\eqref{eq:Ranocha_Flux_Etot} we note that the convective flux for total energy (the first two terms) is split into two contributions, one associated with kinetic energy and the other with internal energy. 
The kinetic energy contribution to the total energy flux is precisely that induced by the KEP discretization of the mass and momentum fluxes. Moreover, the pressure flux $\overline{\overline{\left(p,u\right)}}$ corresponds, in FD terms, with a discretization of the advective form of the conservative pressure term: $\partial_xpu = p\partial_xu+u\partial_xp$ 
with second-order, central schemes~\cite{Coppola2019}. 
This indicates that a formulation based on the internal energy equation, in place of total energy equation, with the conservative convective term specified by the flux $\mathcal{F}_{\rho}\,\left[\overline{\left(1/e\right)}^{\text{log}}\right]^{-1}$ and the pressure term discretized as $p\partial_xu$ with central schemes, is equivalent, for exact time integration, to the formulation expressed in Eq.~\eqref{eq:Ranocha_Flux_Etot} (cf.~\cite{DeMichele2023} for a more general discussion).
To facilitate the comparison of the Ranocha flux with other fluxes to be discussed, from now on we will adopt the expression for this flux in terms of a discretization of mass, momentum and internal energy, for which the convective part of the fluxes is
\begin{equation}\label{eq:Ranocha_Flux_eint}
\mathcal{F}_{\rho}^c= \overline{\rho}^{\text{log}}\,\overline{u},\qquad\qquad
\mathcal{F}_{\rho u}^c =\mathcal{F}_{\rho}^c\,\overline{u},\qquad\qquad
\mathcal{F}_{\rho e}^c = \mathcal{F}_{\rho}^c\,\left[\overline{\left(1/e\right)}^{\text{log}}\right]^{-1}.
\end{equation}

In the framework of FD conservative and KEP formulations, corresponding to fluxes adopting bilinear or trilinear interpolations, there are some that, although not exactly EC, show robust behavior and remarkably good entropy-conservation properties. A quite comprehensive analysis of the various possible approaches has been recently presented~\cite{DeMichele2023}. Here we select two of the most robust ones; the first one is formulated by assuming the same KEP flux for momentum and internal energy
\begin{equation}\label{eq:Flux_eint}
\mathcal{F}_{\rho}^c= \overline{\rho}\,\overline{u},\qquad\qquad
\mathcal{F}_{\rho u}^c =\mathcal{F}_{\rho}^c\,\overline{u},\qquad\qquad
\mathcal{F}_{\rho e}^c = \mathcal{F}_{\rho}^c\overline{e}.
\end{equation}
This flux has been already presented in the literature~\cite{Coppola2019,DeMichele2023,Kuya2018}
and corresponds with a direct discretization of the internal energy equation in which a fully `triple' splitting of the derivative of the product $\rho ue$ is discretized~\cite{Kennedy2008,Coppola2019}. Given exact time integration, this method can be equivalently reformulated as a set of fluxes for mass, momentum and total energy, as in Eq.~\eqref{eq:Ranocha_Flux_Etot} and is equivalent to the KEEP schemes proposed by Kuya \emph{et al.}~\cite{Kuya2018}.

The second approach employs the fluxes
\begin{equation}\label{eq:Flux_soundspeed}
\mathcal{F}_{\rho}^c= \overline{\rho}\,\overline{u},\qquad\qquad
\mathcal{F}_{\rho u}^c =\mathcal{F}_{\rho}^c\,\overline{u},\qquad\qquad
\mathcal{F}_{\rho e}^c = \mathcal{F}_{\rho}^c\overline{e}^G
\end{equation}
where $\overline{\phi}^G=\sqrt{\phi_i\phi_{i+1}}$ is the geometric mean.
This method differs from the previous one only with respect to the internal energy convective flux, which is calculated as the product between the mass flux and the geometric mean for $e$, in place of the arithmetic mean.  
To understand the origin of this formulation, we observe that the form of $\mathcal{F}^c_{\rho e}$ in Eq.~\eqref{eq:Flux_soundspeed} has a strong analogy with the induced kinetic-energy convective flux in a KEP formulation $\mathcal{F}_{\rho u^2/2}^c=\mathcal{F}_{\rho}^cu_iu_{i+1}/2=\mathcal{F}_{\rho}^c\overline{u^2}^G/2$ (we implicitly assume $u_iu_{i+1}\geq 0$). This suggests that the method in Eq.~\eqref{eq:Flux_soundspeed} can be equivalently reformulated as a discretization of the evolution equation for $\rho\sqrt{e}$ (which is proportional to sound speed) with a KEP scheme ensuring a conservative structure of the convective terms in the induced equation for $\rho e$, with a flux analogous to that of kinetic energy in Eq.~\eqref{eq:Ranocha_Flux_Etot}. 
Also this approach has been investigated in previous papers~\cite{DeMichele2023,DeMichele2022b} and is inspired by the work of Kok~\cite{Kok2009}, in which the sound speed equation is used with a different specification of the mass flux.

The methods in Eq.~\eqref{eq:Flux_eint} and \eqref{eq:Flux_soundspeed} can be seen as two dual methods based on the discretization of the internal energy equation. In Eq.~\eqref{eq:Flux_eint} a KEP formulation is used in such a way that the convective term in the induced equation for $\rho e^2$ is in conservation form, whereas in Eq.~\eqref{eq:Flux_soundspeed} one has a formulation in which the discrete balance equation for $\rho\sqrt{e}$ has a convective term in conservation form.
We stress here again that the two methods in Eq.~\eqref{eq:Flux_eint} and \eqref{eq:Flux_soundspeed} can be equivalently expressed as conservative methods for mass, momentum and total energy by using the same corrections needed to pass from the method in Eq.~\eqref{eq:Ranocha_Flux_eint} to its original version in Eq.~\eqref{eq:Ranocha_Flux_Etot}.

\section{Relation among the formulations}
By looking at the formulations in Eq.~\eqref{eq:Ranocha_Flux_eint}, \eqref{eq:Flux_eint} and \eqref{eq:Flux_soundspeed}, we see that the three methods differ for the mean value adopted for $\rho$ in the mass flux ($\overline{\rho}^{\text{log}}$ in the  method of Eq.~\eqref{eq:Ranocha_Flux_eint} and $\overline{\rho}$ in the other two methods) and for that of $e$ in the internal energy flux ($\left[\overline{1/e}^{\text{log}}\right]^{-1}$ in the method of Eq.~\eqref{eq:Ranocha_Flux_eint} and $\overline{e}$ or $\overline{e}^G$ in the methods of Eq.~\eqref{eq:Flux_eint} and \eqref{eq:Flux_soundspeed}, respectively).

The arithmetic, geometric and logarithmic means are related by a classical inequality, which is valid for positive quantities~(cf.~\cite{Carlson1966}, Eq.~(3.1))
\begin{equation}\label{eq:MeansInequality1}
    \gmean{\phi}\leq\logmean{\phi}\leq\mean{\phi}
\end{equation}
with strict inequality for $\phi_i\neq \phi_{i+1}$. 
Since the following simple exact relation between $\gmean{\phi}$ and $\mean{\phi}$ is valid
\begin{equation}\label{eq:GAmeans_relation}
    \gmean{\phi} = \mean{\phi}\left(1-\left(\dfrac{\delta \phi}{2\mean{\phi}}\right)^2\right)^{1/2}
\end{equation}
with $\delta \phi = \phi_{i+1}-\phi_{i}$, it is readily seen that for small non-dimensional values of $\delta\phi$ the arithmetic and geometric means are increasingly good approximations of the logarithmic mean.
These considerations justify, at a qualitative level, the good performances of the methods based on the arithmetic and geometric means.
In the next paragraph, we will analyze this relation from a more quantitative point of view.

In the flux for internal energy also the reciprocal of the logarithmic mean of $e^{-1}$ appears. By applying Eq.~\eqref{eq:MeansInequality1} to the reciprocal of $\phi$ and by using the fact that the geometric mean of the reciprocal equals the reciprocal of the geometric mean, one can easily also show the inequality
\begin{equation}\label{eq:MeansInequality2}
\hmean{\phi}\leq\left(\logmean{\left(1/\phi\right)}\right)^{-1}\leq\gmean{\phi}
\end{equation}
where $\hmean{\phi}=\phi_i\phi_{i+1}/\mean{\phi}$ is the \emph{harmonic} mean.
Moreover, it is a simple exercise to show that from Eq.~\eqref{eq:GAmeans_relation} also the exact relation between $\hmean{\phi}$ and $\gmean{\phi}$ follows
\begin{equation}\label{eq:HGmeans_relation}
    \hmean{\phi} = \gmean{\phi}\left(1-\left(\dfrac{\delta \phi}{2\mean{\phi}}\right)^2\right)^{1/2}.
\end{equation}
Eqs.~\eqref{eq:MeansInequality1}-\eqref{eq:HGmeans_relation} make it possible to draw some important conclusions.
The first one is that the geometric mean stands as a good approximation for both the logarithmic mean of the density in the mass flux and for the average based on the logarithmic mean  appearing in the internal energy flux. In fact, a formulation based on the geometric mean for both $\rho$ and $e$
\begin{equation}\label{eq:Flux_geomgeom}
\mathcal{F}_{\rho}^c= \gmean{\rho}\,\overline{u},\qquad\qquad
\mathcal{F}_{\rho u}^c =\mathcal{F}_{\rho}^c\,\overline{u},\qquad\qquad
\mathcal{F}_{\rho e}^c = \mathcal{F}_{\rho}^c\overline{e}^G,
\end{equation}
which is consistent with the formulation studied in~\cite{Rozema2014},
shows remarkably good properties in terms of entropy conservation for our test cases (cf.~Sec.~\ref{sec:Results}).
The second conclusion is that, while the arithmetic mean is fully justified for the approximation of the logarithmic mean of the density in the mass flux by virtue of Eq.~\eqref{eq:MeansInequality1}, its use for the internal energy in the internal energy flux seems less obvious. 
Eq.~\eqref{eq:MeansInequality2} suggests that either the harmonic mean or the geometric mean is a more legitimate candidate to be a better approximation of the internal energy average.
Since the original motivation of the present analysis is to design a flux that is based only on algebraic operations, from now on we will focus on a formulation involving the arithmetic mean for density and the harmonic mean for internal energy
\begin{equation}\label{eq:Flux_ArithHarmonic}
\mathcal{F}_{\rho}^c= \mean{\rho}\,\overline{u},\qquad\qquad
\mathcal{F}_{\rho u}^c =\mathcal{F}_{\rho}^c\,\mean{u},\qquad\qquad
\mathcal{F}_{\rho e}^c = \mathcal{F}_{\rho}^c\hmean{e}.
\end{equation}
In the next section, we show that a correct asymptotic expansion of the logarithmic mean naturally leads to the formulation in Eq.~\eqref{eq:Flux_ArithHarmonic} as a first-order approximation.
Moreover, our treatment gives a class of increasingly accurate fluxes which retain at each finite order of approximation the PEP property, in contrast to the formulations in Eq.~\eqref{eq:Flux_eint} and \eqref{eq:Flux_soundspeed} and to other existing methods based on asymptotic approximations.

\section{Asymptotic expansion of the logarithmic mean}

As previously mentioned, most EC fluxes require the evaluation of logarithmic means which comes with an increased computation cost compared to algebraic fluxes. To overcome this disadvantage, it is possible to expand the logarithmic mean in a Taylor series in the small parameter $\delta \phi$. 
This approach is not entirely new, as it was already used in~\cite{Ismail2009} to resolve the singularity of the logarithmic mean when uniform distribution of $\phi$ appears. 
Starting from a different perspective, a similar formulation is obtained also in~\cite{Tamaki2022}, who 
proposes an asymptotically EC formulation that shares many similarities with the approach used here. However, our formal expansion of the logarithmic mean in density and internal energy fluxes leads to a different formulation, whose first-order approximation involves the harmonic mean for $e$ in place of the arithmetic mean.

We start by expressing the difference of logarithms for a generic quantity $\phi$ as 
\begin{equation}
 \diff \log{\phi_i} =
 \log\left(1 + \hat{\phi_i}\right)
 -\log\left(1 - \hat{\phi_i}\right)
\end{equation}
with $\hat{\phi_i} =  (\diff \phi_i)/(2\mean{\phi_i})$ already appearing in Eq.~\eqref{eq:GAmeans_relation} and 
Eq.~\eqref{eq:HGmeans_relation}. Since the quantity $\lvert\hat{\phi}\rvert$ is always less than one for a positive $\phi$, it is possible to use the Taylor series expansion for the logarithm and obtain
\begin{equation}
\diff \log{\phi} =
 \left(\frac{ \delta \phi}{\mean{\phi}}\right)\sum_{n=0}^\infty \frac{\hat{\phi}^{2n}}{2n+1} .
 \label{eq:log_expansion}
\end{equation} 
Applying this substitution to $\diff \log{\rho}$ and $\diff \log{e}$ in the logarithmic means $\logmean{\rho}$ and $\logmean{e^{-1}} = - (\delta\log e )/(\delta e^{-1})$ in Eq.~\eqref{eq:Ranocha_Flux_eint} and truncating the sum to finite $N$, we obtain the class of AEC fluxes
\begin{equation}\label{eq:Flux_eint_expansion}
\mathcal{F}_{\rho}^c= \mean{\rho}\mean{u}\left(\sum_{n=0}^N \frac{\hat{\rho}^{2n}}{2n+1}\right)^{-1},\qquad\qquad
\mathcal{F}_{\rho u}^c =\mathcal{F}_{\rho}^c\,\mean{u},\qquad\qquad
\mathcal{F}_{\rho e}^c = \mathcal{F}_{\rho}^c\,\hmean{e}\sum_{n=0}^N \frac{\hat{e}^{2n}}{2n+1}.
\end{equation}
Note that this expansion differs from that adopted by \cite{Tamaki2022}, 
since the asymptotic expansion is applied to $\delta \log e$ in the internal energy flux, but not to $\diff e^{-1}$.
In fact, it leads to the appearance of the harmonic mean in place of the arithmetic mean in the internal energy flux.
Eq.~\eqref{eq:Flux_eint_expansion} reduces to Eq.~\eqref{eq:Flux_ArithHarmonic} in the first-order case $N=0$ 
and to Eq.~\eqref{eq:Ranocha_Flux_eint} for $N\to\infty$.
One additional property of the class of schemes in Eq.~\eqref{eq:Flux_eint_expansion} is that it is always PEP, no matter the value of $N$, as long as it is chosen consistently for density and internal energy expansions. 

In order for a scheme to be PEP, for constant $u = U$ and $p = P$, the conditions on the fluxes are $\mathcal{F}_{\rho u} = \mathcal{F}_{\rho}\, U + \text{const}$ (in which the constant is only function of $U$ and $P$) and that $\mathcal{F}_{\rho e}$ should be equal to a constant dependent on only $U$ and $P$~\cite{Ranocha2021}. In these hypotheses, $\hmean{e} = P/[(\mean{\rho}(\gamma-1)]$ and $\hat{e} = -\hat{\rho}$, so
\begin{equation}
\mathcal{F}_{\rho e}^c = \mean{\rho}\hmean{e}U\left(\sum_{n=0}^N \frac{\hat{e}^{2n}}{2n+1}\right)\left({\sum_{n=0}^N \frac{\hat{\rho}^{2n}}{2n+1}}\right)^{-1} = \frac{UP}{\gamma -1}
\end{equation}
which proves the PEP property for the fluxes in Eq.~\eqref{eq:Flux_eint_expansion}.
In an analogous way, it can be shown that also the fluxes in Eq.~\eqref{eq:Flux_geomgeom} produce a PEP scheme.
\section{Numerical results}\label{sec:Results}

Two tests have been performed to assess the properties of the various schemes. The first one is a density wave test analogous to those in \cite{Shima2021,Ranocha2021}, which is useful to ascertain the PEP property since $u$ and $p$ are initially constant;
the second one is the inviscid Taylor-Green vortex, which 
has been used to test the higher-order fluxes in a three-dimensional case
in which an initially smooth flow experiences distortion and instability, with the eventual formation of small unresolved scales. 
For the first test, a finite volume method has been used with second-order accurate fluxes for the primary variables $\rho,\rho u$ and $\rho e$; the pressure terms have been discretized using second-order central schemes.
For the Taylor-Green vortex, both fourth and sixth-order accurate fluxes of $\rho,\rho u$ and $\rho E$ have been used, 
as defined by Eq.~\eqref{eq:high_order_flux}.
The classical fourth-order Runge-Kutta (RK4) has been employed with $\textrm{CFL} = 0.01$ in the first test and with $\textrm{CFL} = 0.1$ in the second one.

In this section, the results obtained using six different methods will be presented. They will be identified by using a string indicating the mean adopted and the corresponding variable in mass and internal energy fluxes: $A \rho$-$A e$ is used for the fluxes in Eq.~\eqref{eq:Flux_eint}, which is equivalent to KGP($\rho e$) in \cite{Coppola2019} and---for exact time integration---to KEEP in \cite{Kuya2018}; $A \rho$-$H e$ identifies the new method with fluxes from Eq.~\eqref{eq:Flux_ArithHarmonic};
$G \rho$-$G e$ describes the fluxes using the geometric mean in Eq.~\eqref{eq:Flux_geomgeom}; 
$A \rho$-$A p$ identifies a formulation in which the arithmetic mean is used in the internal energy equation by grouping 
$\rho$ from the mass flux and $e$:
\begin{equation}\label{eq:Flux_Arith_rho_rhoe}
\mathcal{F}_{\rho}^c= \mean{\rho}\,\overline{u},\qquad\qquad
\mathcal{F}_{\rho u}^c =\mathcal{F}_{\rho}^c\,\mean{u},\qquad\qquad
\mathcal{F}_{\rho e}^c = \mean{u}\mean{\rho e} = \dfrac{1}{\gamma-1}\mean{u}\mean{p},
\end{equation}
which corresponds to the scheme KEEP-PE in \cite{Shima2021} and is an exemplary PEP scheme.
Finally, AEC${}^{(1)}$ uses the fluxes in Eq.~\eqref{eq:Flux_eint_expansion} with $N=1$, whereas KEEP${}^{(1)}$ refers to the formulation of \cite{Tamaki2022}, which is similar to that reported in Eq.~\eqref{eq:Flux_eint_expansion}, except for the internal energy flux, given in our notation by $\mathcal{F}_{\rho e}^c = \mathcal{F}_{\rho}^c\,\mean{e}(1 + \hat{e}^{2}/{3})/(1 + \hat{e}^{2})$.
The method using the fluxes in Eq.~\eqref{eq:Ranocha_Flux_eint} has also been tested; the results concerning entropy production are not shown since it is always equal to machine zero.

\begin{figure}[tb]
    \centering
    \begin{subfigure}[b]{0.47\textwidth}
         \centering
         \includegraphics[width=\textwidth]{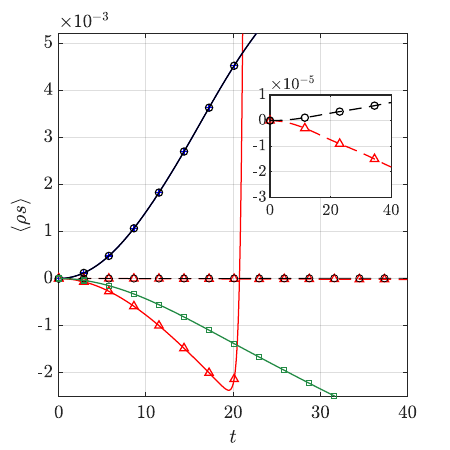}
         \caption{Entropy integral time evolution}
         \label{fig:entropy_integral_DW}
     \end{subfigure}
    \begin{subfigure}[b]{0.47\textwidth}
         \centering
         \includegraphics[width=\textwidth]{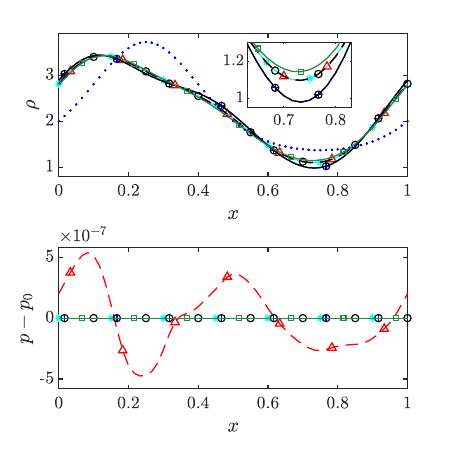}
         \caption{Solution comparison}
         \label{fig:solution_DW}
     \end{subfigure}
    \caption{Density wave simulation using different numerical fluxes. On the left, time evolution of entropy integral:
    black continuous lines with circles represent the $A\rho$-$He$ scheme; red continuous lines with triangles represent the $A\rho$-$Ae$ scheme; blue with plus signs identifies the $A\rho$-$Ap$ scheme; green with squares is used for the geometric mean flux $G\rho$-$Ge$. 
    Dashed lines represent the AEC${}^{(1)}$ (black and circles) and KEEP${}^{(1)}$ (red and triangles) schemes. On the right, comparison of the density and pressure at time $t=30$; the exact solution is represented by a blue dotted line, the cyan line with asterisk markers is the solution obtained using the Ranocha flux in Eq.~\eqref{eq:Ranocha_Flux_eint}; the solution using the $A\rho$-$Ae$ scheme is not shown, as the simulation had already diverged.
    The mesh is discretized in 61 nodes
    and $\textrm{CFL} = 0.01$.}
    \label{fig:entropy_integral}
\end{figure}
The initial conditions for the density wave test are
\begin{equation*}
    \rho_0 = 1 + \exp\left(\sin\left(\frac{2\pi x}{L}\right)\right), \qquad u_0 = 1, \qquad p_0 = 1,
\end{equation*}
with the domain of  size $L = 1$ discretized in 61 points and periodic boundary conditions.
For this test, a more complex initial condition has been chosen for density when compared to the more usual monochromatic sinusoidal wave employed in~\cite{Shima2021,Ranocha2021}. That is because in some cases we experienced fortuitous global conservation of entropy due to the symmetry of the sine function, despite the lack of exact local conservation.
Figure~\ref{fig:entropy_integral_DW} shows the temporal evolution of the quantity $\langle \rho s\rangle$, which is the normalized global entropy production $(\widetilde{\rho s} - \widetilde{\rho_0 s_0} )/(\widetilde{\rho_0 s_0})$, with $\rho_0 s_0$ being the initial value and the $\ \widetilde{}\ $ sign indicating integration over the domain. 
The $A\rho$-$A e$ scheme, which is not PEP, diverges around time $t=22$, whereas the $A\rho$-$H e$ and $G\rho$-$G e$ schemes
show enhanced robustness; although the entropy-production curves for these schemes show an important growth for the scales of the plot, they remain stable in our simulations, which have been extended up to $t=100$. Note that the entropy production of the $A\rho$-$A p$ scheme and that of the $A\rho$-$H e$ are identical as it can be predicted theoretically: in fact, both methods use the same flux for the density equation and, for constant $u$ and $p$, the convective flux of internal energy is $\mathcal{F}_{\rho e}^c = UP/\left(\gamma -1\right)$ for both.
This implies that until pressure and velocity remain constant, as in the present test, the two formulations behave identically, up to round-off error.
Fig.~\ref{fig:solution_DW} (upper panel) reports the numerical solution for all the schemes analyzed at $t=30$, at which the error on entropy reaches the value $\langle \rho s \rangle = 7\times 10^{-3}$ for $A\rho$-$H e$ and $A\rho$-$Ap$ schemes and $-2\times 10^{-3}$ for $G\rho$-$G e$.
Since all the schemes tested are only second-order accurate, noticeable numerical dispersion is visible comparing the density results to the exact solution (Fig.~\ref{fig:solution_DW}, upper panel).
KEEP$^{(1)}$ and AEP$^{(1)}$, which have better entropy-conservation properties, behave similarly for this test, with a slightly lower entropy production for the latter, except that KEEP$^{(1)}$ presents spurious pressure and velocity oscillations due to the lack of the PEP property, as shown in Fig.~\ref{fig:solution_DW}, (lower panel).
 Again, $A\rho$-$A p$ and $A\rho$-$H e$ perform identically for this test due to their equivalence, whereas KEEP$^{(1)}$ and AEP$^{(1)}$ produce results for density which are visually indistinguishable from that obtained using the flux in Eq.~\eqref{eq:Ranocha_Flux_eint}. The results of $A\rho$-$A e$ are not shown in Fig.~\ref{fig:solution_DW} due to the earlier blow-up. 

\begin{figure}[tb]
    \centering
    \begin{subfigure}[b]{0.47\textwidth}
         \centering
         \includegraphics[width=\textwidth]{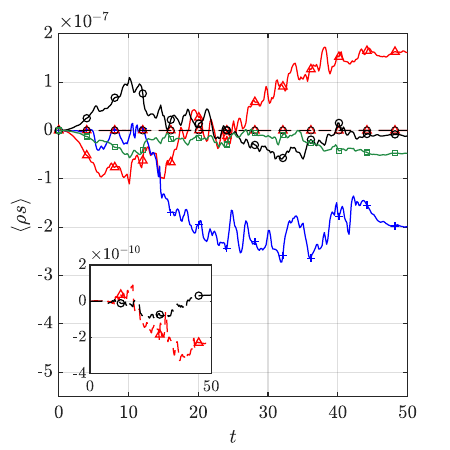}
         \caption{Fourth-order accurate fluxes}
         \label{fig:entropy_integral_TGV_IV}
     \end{subfigure}
    \begin{subfigure}[b]{0.47\textwidth}
         \centering
         \includegraphics[width=\textwidth]{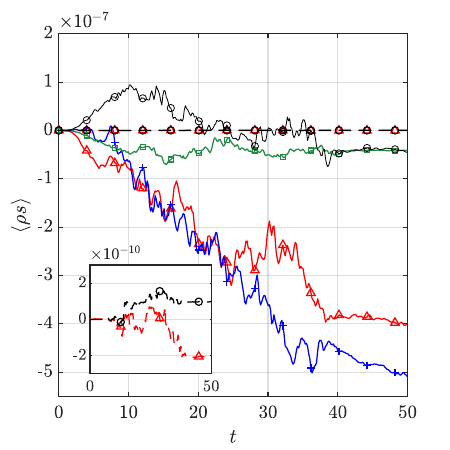}
         \caption{Sixth-order accurate fluxes}
         \label{fig:entropy_integral_TGV_VI}
     \end{subfigure}
    \caption{Time evolution of entropy integral for the inviscid Taylor-Green vortex test using different numerical fluxes: black continuous lines with circles represent the $A\rho$-$He$ scheme; red continuous lines with triangles represent the $A\rho$-$Ae$ scheme; blue with plus signs identifies the $A\rho$-$Ap$ scheme; green with squares is used for the geometric mean flux $G\rho$-$Ge$. 
    Dashed lines represent the AEC${}^{(1)}$ (black with circles) and KEEP${}^{(1)}$ (red with triangles) schemes. The mesh is discretized using $32\times32\times32$ nodes;  fourth-order accurate fluxes are used on the left figure, sixth-order is employed for the figure on the right. In both cases $\textrm{CFL} = 0.1$.}
    \label{fig:entropy_integral_TGV}
\end{figure}

The initial conditions for the Taylor-Green vortex are
\begin{align*}
    \rho(x,y,z) &= 1\\
    u(x,y,z) &= \sin(x)\cos(y)\cos(z)\\
    v(x,y,z) &= -\cos(x)\sin(y)\cos(z)\\
    w(x,y,z) &= 0\\
    p(x,y,z) &= 10 + \frac{(\cos(2x) + \cos(2y))(\cos(2x)+2)-2}{16}
\end{align*}
 with a pressure value corresponding to a Mach number $M \approx 0.26$.
The triperiodic domain has side length $2\pi$ in all directions and is discretized using $32\times32\times32$ nodes. The chosen CFL value is sufficiently small that linear invariants are exactly conserved to machine precision for all schemes. The time evolution of the entropy integral for this test is shown in Fig.~\ref{fig:entropy_integral_TGV} and it is in agreement with the previous results.
In this test, since the pressure is not constant, $A\rho$-$H e$ is no longer equivalent to $A\rho$-$A p$;
in this case we have better performances from $A\rho$-$H e$ and $G\rho$-$G e$ when compared to $A\rho$-$A p$ and $A\rho$-$A e$ and this result is found for both fourth-order and six-order accurate fluxes. An improvement can be obtained using an additional term in the expansions and KEEP$^{(1)}$ and AEP$^{(1)}$ are the schemes which more closely achieve a constant value for the entropy integral.
Information about the reliability of the scheme can be obtained thorough the study of the evolution of thermodynamic fluctuations in time. 
We checked that for all the schemes tested, the density and temperature fluctuations do not have an unbound growth (not shown). This is the desired behavior, since for inviscid isotropic homogeneous turbulence they are reported to level off to a constant value \cite{HoneinMoin2004,Coppola2019}.

\section{Conclusions}

We proposed a new class of asymptotically entropy-preserving fluxes for the discretization of the convective terms in the compressible Euler equations with interesting properties. 
It provides a consistent asymptotic approximation of an existing entropy-preserving scheme based on the logarithmic mean, and it consists of economical algebraic fluxes based on the harmonic mean. Moreover, at all orders of approximation, the numerical fluxes have the pressure-equilibrium preservation property.
The theoretical predictions are confirmed on two test cases, verifying that the new schemes are able to numerically maintain pressure equilibrium and demonstrating good entropy-conservation property. It was also shown that the error on entropy can be reduced by using additional terms in the expansion of the AEC fluxes.

These results suggest that AEC fluxes could be good candidate for the discretization of compressible flow equations in high performance solvers.
Due to their algebraic form, they are less computationally expensive than the fluxes based on the logarithmic mean, while retaining many important properties. In fact, they guarantee the KEP and PEP properties, combined with arbitrarily small error on entropy preservation.
\appendix
\section{High-order extension}
\label{sec:high_order}
The second-order accurate two-point fluxes presented in this article can be extended to higher-order formulations by using the approach proposed by Ranocha~\cite{Ranocha2018} in the context of Discontinuous Galerkin discretization of the Euler equations.
The main result of interest for us is that contained in Theorem 3.1 of \cite{Ranocha2018}, which can be reformulated in  FD terms as follows.
We consider a numerical flux $\mathcal{F}(\vec{w_i},\vec{w_{i+k}})$ for a generic quantity $\rho\phi$, which depends on the values of the variables vector $\vec{w}$ in the nodal points $i$ and $i+k$.
In our context $\mathcal{F}$ can be any of the numerical fluxes specified in 
Eqs.~\eqref{eq:Ranocha_Flux_Etot}--\eqref{eq:Flux_soundspeed},\eqref{eq:Flux_geomgeom}--\eqref{eq:Flux_ArithHarmonic} or
\eqref{eq:Flux_eint_expansion}--\eqref{eq:Flux_Arith_rho_rhoe}
and $\vec{w}$ is the set of variables $\left(\rho,u,e\right)$. 
We will assume that the numerical flux is smooth, symmetrical 
(i.e. $\mathcal{F}(\vec{w_i},\vec{w_{i+k}}) = \mathcal{F}(\vec{w_{i+k},\vec{w_i}})$) and consistent with the continuous flux $f$ so that $\mathcal{F}(\vec{w_i},\vec{w_{i}}) = f(\vec{w_i})$.
Under these hypotheses, by following the steps of the proof to Theorem~3.1 in~\cite{Ranocha2018}, we can show that  given a numerical derivative formula of the type $\partial\varphi_i\simeq\sum_{k}a_{k}\varphi_{i+k}$
then an approximation of the derivative $\partial f$ is given by
\begin{equation}\label{eq:high_order_ranocha}
    \sum_{k} 2 a_{k} \mathcal{F}(\vec{w_i},\vec{w_{i+k}})
\end{equation}
and it has the same order of accuracy as the original derivative formula with weights $a_{k}$.
If one considers central derivative formulas, for which $a_{k} = -a_{-k}$, by using the symmetry of the flux $\mathcal{F}$, Eq.~\eqref{eq:high_order_ranocha} can be rewritten as
\begin{equation}\label{eq:AppEq2}
        \sum_{k=1}^{L} 2 a_{k}
    \left(
    \mathcal{F}(\vec{w_i},\vec{w_{i+k}})- \mathcal{F}(\vec{w_{i-k}},\vec{w_i}) 
    \right).
\end{equation}
Adding and subtracting the terms $\sum_{m=1}^{k-1}\mathcal{F}(\vec{w_{i-m}},\vec{w_{i-m+k}})$
as in~\cite{Pirozzoli2010}, one can recast Eq.~\eqref{eq:AppEq2} as the difference of high-order numerical fluxes $\mathcal{F}_{i+1/2}^{h}-\mathcal{F}_{i-1/2}^{h}$ where
\begin{equation}\label{eq:high_order_flux}
    \mathcal{F}_{i+1/2}^{h} = 2\sum_{k=1}^{L} a_{k}
    \sum_{m=0}^{k-1}\mathcal{F}(\vec{w_{i-m}},\vec{w_{i-m+k}}),
\end{equation}
which is the high-order extension of the second-order, two-point fluxes illustrated in the present paper.
The validity of this result is already well established in the simple cases in which the numerical fluxes 
are built by using only arithmetic or product averages, as for example in Eq.~\eqref{eq:Flux_eint} or \eqref{eq:Flux_Arith_rho_rhoe}.
In these cases, Eq.~\eqref{eq:high_order_flux} reduces
to a FD discretization of a linear combination of advective and divergence forms of the convective terms, 
discretized with a high-order numerical derivative 
formula (cf.~Eqs.~(A.5)--(A.9) in \cite{Coppola2019}
or Eqs.~(20)--(21) in \cite{DeMichele2023}).  
Eq.~\eqref{eq:high_order_flux} extends this result to arbitrary nonlinear fluxes, for which a
formulation expressed through a direct discretization of divergence and advective forms does not exist in general.

\bibliographystyle{unsrtnat}
\bibliography{references} 

\end{document}